\documentclass[twocolumn,showpacs,preprintnumbers,amsmath,amssymb,superscriptaddress]{revtex4}
\usepackage{graphicx}
\usepackage{dcolumn}
\usepackage{bm}

\begin{document}
\preprint{preprint}

\title{Muon spin rotation and relaxation studies on the solid solution of the 
two spin-gap systems (CH$_3$)$_2$CHNH$_3$-CuCl$_3$ and (CH$_3$)$_2$CHNH$_3$-CuBr$_3$
}
\author{T. Saito} 
\author{A. Oosawa} 
\author{T. Goto} 
\email{gotoo-t@@sophia.ac.jp}
\affiliation{Department of Physics, Sophia University, 7-1 Kioi-cho, Chiyoda-ku, Tokyo 102-8554, Japan}
\author{T. Suzuki}
\affiliation{Advanced Meson Science Laboratory, RIKEN 
(The Institute of Physical and Chemical Research)\\
2-1 Hirosawa, Wako, Saitama 351-0198}
\affiliation{Department of Physics, Sophia University, 7-1 Kioi-cho, Chiyoda-ku, Tokyo 102-8554, Japan}
\author{I. Watanabe}
\affiliation{Advanced Meson Science Laboratory, RIKEN 
(The Institute of Physical and Chemical Research)\\
2-1 Hirosawa, Wako, Saitama 351-0198}

\date{\today}

\begin{abstract}
Muon-spin-rotation and relaxation studies have been performed on (CH$_3$)$_2$CHNH$_3$Cu(Cl$_x$Br$_{1-x}$)$_3$ 
with $x$=0.85 and 0.95, 
which are solid solutions of the two isomorphic spin-gap systems 
(CH$_3$)$_2$CHNH$_3$CuCl$_3$ and (CH$_3$)$_2$CHNH$_3$CuBr$_3$ with different spin gaps.   
The sample with $x$=0.85 showed a clear muon spin rotation under zero-field 
below $T_{\rm N}$=11.65K, 
indicating the existence of a long-range antiferromagnetic order.    
A critical exponent of the hyperfine field was obtained to be $\beta$=0.33, which 
agrees with 3D-Ising model.
In the other sample with $x$=0.95, an anomalous enhancement of 
the muon spin relaxation was observed at very low temperatures indicating a 
critical slowing down due to a magnetic instability of the ground state.
\end{abstract}

\maketitle

\section{Introduction}
Antiferromagnetically interacting spin systems with a low dimensional structure often show a ground state
unexpected from a classical picture.   
This is due to a quantum spin fluctuation, which destabilizes the N$\acute{\rm e}$el state 
and tends to bring the system to the non-magnetic state such as the Haldane state 
or the singlet dime state.  
The title compounds (CH$_3$)$_2$CHNH$_3$-CuCl$_3$ and (CH$_3$)$_2$CHNH$_3$-CuBr$_3$ 
abbreviated as IPA-CuCl$_3$ and IPA-CuBr$_3$ are low-dimensional spin-gap systems with an isomorphic crystal 
structure~\cite{1st_report_IPA_Cl3_chi_T,chi-T_in_Br_system}.
The origin of the spin gap is quite different for these two compounds.   
In Cl-system, which has recently been revealed by an inelastic neutron experiment
to be a spin ladder\cite{Chain_and_ladder_Masuda},
the most dominant coupling on the neighboring two spins with $S$=$\frac{1}{2}$ on 
the rung leads a formation of a ferromagnetic dimer.   
These fake $S$=1 spins interact weakly with antiferromagnetic bonds along a ladder, 
so that the system is well considered as the Haldane system, that is, 
$S$=1 antiferromagnetic 
Heisenberg chain~\cite{1st_report_IPA_Cl3_chi_T,Haldane_original,ESR_D_term,ESR_D_term2}.
The spin excitation gap in this system is reported to be around 10K~\cite{M_H_on_Cl_system}.   
While in Br-system, the dominant interaction between the two neighboring spins is antiferromagnetic, 
so that at low temperatures, they form a non magnetic dimer, which interacts weakly 
with other dimers.
The spin excitation gap is reported to be around 80-100K~\cite{chi-T_in_Br_system, is_Br_ladder_or_not}.
Though these two Cl and Br systems look quite different,
one should notice that the underlying physics is the same in that 
the two have a gapped ground state.

The subject of this study is to investigate what will happen when these two non magnetic
systems with quite different spin-spin couplings are mixed.
In a classical picture, one should note that a uniform antiferromagnetic ordering is instable 
against bond randomness, and that a spin-glass state is favorable.   
This is because ferromagnetic bonds embedded randomly in antiferromagnetic bonds randomly reverse 
the staggered phase factor of the antiferromagnetism, and hence destabilize it.

In the quantum mechanical picture, however, the effect of randomness is not self-evident.   
Disorder usually breaks the order, but sometimes it brings order to disorder\cite{Nakamura_theory}.
For example, it is well known that doped holes of only three percent destroy the N$\acute{\rm e}$el
 order in high-$T_{\rm C}$ cuprates~\cite{Kitaoka_LBCO_La_NQR,Review_on_HTSC_mainly_on_neutron}.
This is explained in terms of the frustration effect.   
In spin Peierls systems, on the contrary, 
an extremely small fraction of non magnetic impurities causes the antiferromagnetic order 
by destructing singlet 
dimers~\cite{CuGeO3_Zn_dope_1st_Hase,CuGeO3_Zn_dope_detailed_phase_diagram}.   
In the present system, where there is neither a frustration effect nor 
a destruction of a dimer, we expect a novel-type impurity effect.

So far, investigations on macroscopic quantities such as uniform magnetization and 
specific heat on 
IPA-Cu(Cl$_{x}$Br$_{1-x}$)$_3$ have shown 
that there exists a phase transition in the limited range of 
the mixing ratio $x$=0.44-0.87~\cite{chi_and_C_on_mixed_system_Br_Cl,M_H_on_mixed_system_Br_Cl}.   
From a uniaxial anisotropy in the magnetization observed below a certain temperature, 
they claim that an antiferromagnetic-like ordering takes place at low temperatures.   
The dependence of the transition temperature on the mixing ratio 
indicates that there exist the two quantum critical points, and that both the 
two are of the first order.

As a theoretical approach, Nakamura has studied the 
$S$=$\frac{1}{2}$ ferromagnetic and antiferromagnetic random alternating Heisenberg 
quantum spin chain by a 
quantum Monte-Carlo method to show that a uniform antiferromagnetic 
state is stable in a certain region of $x$, and that there exists a quantum phase transition 
at the edge of the region~\cite{Nakamura_theory}.
He also claims that the quantum effect is indispensable to the occurrence of the uniform antiferromagnetic ordering.   
On the other hand, 
Fisher and his colleagues had worked on the lattice boson system by introducing a randomness to the 
onsite chemical potential, and predicted 
the appearance of the new ground state, Bose-glass phase at the absolute zero~\cite{Fisher_Bose_glass}.   
It has been well known that the field-induced or thermally-excited triplets in spin gap systems including 
the present compound can be treated as nearly-free bosons~\cite{BEC_TlCuCl3}, 
and that their theory is believed to be applicable to disordered spin gap systems.   
In fact, their prediction is supported by recent experiments on a solid solution of 
other two spin gap systems TlCuCl$_3$ and KCuCl$_3$, on which a possibility of 
the existence of the Bose-glass phase has 
been reported\cite{TlK_Oosawa_chi,TlK_Shindo_C,TlK_Suzuki_muSR}.

In this paper, we report by a microscopic probe of $\mu$SR 
an extensive study on the spin state in the two samples with mixing ratios $x$=0.80 and 0.95, 
which are in and out of 
the ordering region~\cite{chi_and_C_on_mixed_system_Br_Cl,M_H_on_mixed_system_Br_Cl}.
We expect that this experiment will be of help for understanding the quantum effect of 
disorder in spin gap systems.

\section{Experimental}

Single crystals of (CH$_3$)$_2$CHNH$_3$Cu(Cl$_x$Br$_{1-x}$)$_3$ with $x$=0.85 and 0.95 have been grown by an evaporation 
method from isopropylalcohol solution of (CH$_3$)$_2$CHNH$_2\cdot$HX and CuX$_2$ with 
X=Cl, Br.\cite{1st_report_IPA_Cl3_chi_T,chi-T_in_Br_system}
A mixing ratio $x$ in crystals is determined by the ICP method.    
The outer shape of crystals is a rectangular-solid.   
Following Manaka {\em et al.}, we refer the orthogonal three planes on the outer shape of crystals as 
A, B, and C-plane~\cite{1st_report_IPA_Cl3_chi_T,Masuda_crystal_axes}.

Before the muon experiments, macroscopic quantities of the single crystals were measured 
to confirm their quality and homogeneity.   
The specific heat of the sample $x$=0.85 measured by the thermal relaxation method shows 
a cusp at $T$$\simeq$12K, which agrees with Manaka's report.
The uniform magnetization also shows a significant anisotropy below this temperature, 
suggesting an existence of a phase transition.
No spurious Curie-term is observed at low temperatures down to 2K in either the two samples.   
These observations are quantitatively consistent with 
Manaka's results~\cite{chi_and_C_on_mixed_system_Br_Cl,M_H_on_mixed_system_Br_Cl}. 

Measurements of $\mu$SR have been carried out at Riken-RAL Muon Facility in U. K. 
using a spin-polarized pulsed surface-muon ($\mu^+$) with a momentum of 27MeV/$c$.   
The two single crystals with $x$=0.85 and one with $x$=0.95 were used for muon measurements.   
They are attached by an Apiezon N grease on the silver plate with a purity of four nines, 
connected at the bottom of the cryostat.   
A typical size of crystals are 10$\times$2$\times$20mm$^3$.   
The incident muon beam is perpendicular to C-plane of the crystals.   
The sample temperature was controlled by Oxford $^3$He cryostat in the range of 0.3-14K.   
The typical temperature stability during measurements is $\pm$10mK above 1.5K, 
and $\pm$2mK below 1.5K.
 
\section{Results}
Figure 1 shows the profile of ZF (zero-field) muon rotation spectra of the sample $x$=0.85.   A clear rotation is observed at low temperatures.   
A short time region within 1$\mu$sec of relaxation curves is fit to the function 
$\exp(-\lambda t)\cos(2\pi \nu t+\phi)$, 
where $\lambda$, $\nu$, and $\phi$ are fitting parameters, 
corresponding to an inhomogeneity of the local field, 
a mean value of the local field, and a phase constant.   
The temperature dependence of $\nu$ is shown at Fig. 2, 
where a scaling function $(1-T/T_N)^\beta$ is 
fitted to the data to obtain the transition temperature 
$T_{\rm N}$=11.65($\pm$0.05)K and the critical exponent $\beta$=0.33($\pm$0.02).   
The local field of the muon site at 0.33K is estimated to be 
$H_{\rm loc}$=$\nu/\gamma_\mu$=360 Oe.
When the rotating frequency becomes close
to the pulse width of the muon beam 70nsec/FWHM, 
the oscillating amplitude is appreciably decreased at low temperatures.   
Assuming the shape of the muon pulse as to be Gaussian, we have calculated 
the convolution and found that the amplitude should decrease to be 65\%, 
which is comparable to the observed value of 50\%.   
This reduction is due to the effect of a finite pulse width of the muon burst 
rather than the change in the electronic spin state.   
The relaxation rate $\lambda$ is around 5-7$\mu$sec$^{-1}$ 
in the ordered state, and shows a slight increase below 6K.
The distribution of the internal field is 
$\delta H_{\rm loc}$=$\lambda/2\pi \gamma_\mu$=130-160Oe.

In Fig. 3, we show relaxation curves under various longitudinal fields (LF) 
up to 10Oe.   
In paramagnetic state at 14K, 
the asymmetry change in the entire ZF curve is approximately 8\%.   
The early part of curves within a few micro seconds shows the Gaussian decay, 
indicating that the muon relaxation is caused by nuclear spins, 
which create a quasi static field around muon sites.   
This is confirmed by the observation that the weak LF of 100Oe completely suppresses the relaxation.

When the temperature is lowered to 11.8K, which is higher than $T_{\rm N}$ only by 0.15K, 
no sign of critical slowing down is observed, that is, the relaxation curve 
is identical to those in the paramagnetic state as seen in Fig. 1.   
Below $T_{\rm N}$, as shown in the lower panel of Fig. 3, 
the initial asymmetry is drastically reduced by the large internal field due 
to the magnetic ordering.   
A very slow component of relaxation persists at the lowest temperature 0.33K.  
This relaxation is considered to be caused by a quasi static nuclear spin fluctuation, 
because it is easily suppressed by a weak LF of 10Oe.   
The ZF-asymmetry change due to this nuclear spin component is 4\%, 
which is almost a half of that in paramagnetic state at 14K. 


\begin{figure}[h]
\includegraphics*[trim=2cm 6cm 1cm 2cm, clip, scale=0.5]{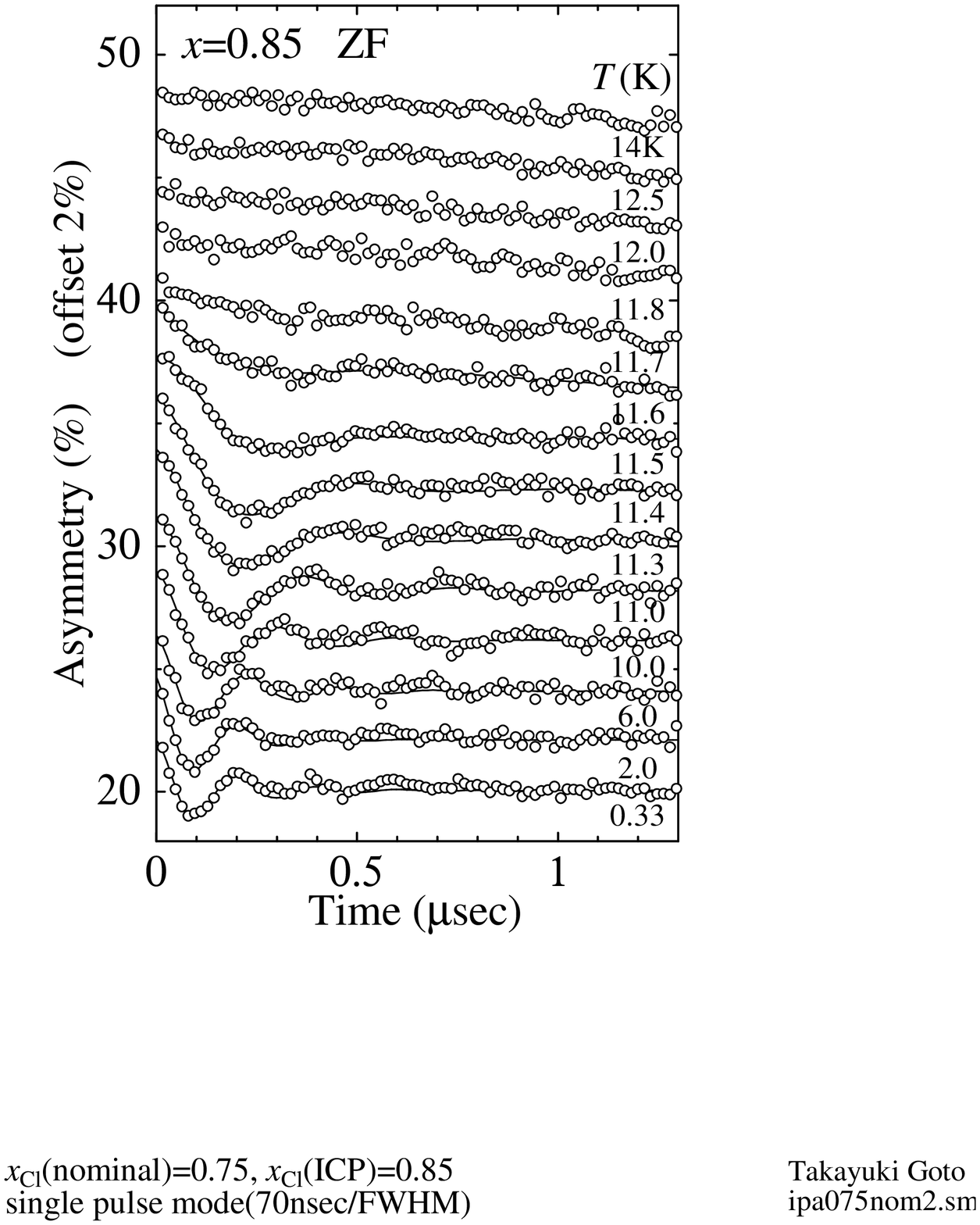}
\caption{
The profile ZF-muon rotating spectra under various temperatures.  
Each spectrum other than that for 0.33K is shifted vertically by 2\%.   
A clear oscillation due to muon rotation was observed at low temperatures below 11.6K.   
Curves are the fitting function 
$e^{-\lambda t}\cos(2\pi\nu t+\phi)$, where $\lambda$, $\nu$, $\phi$ are fitting constants.
}
\end{figure}

Next, we proceed to the results on the sample with $x$=0.95, which shows no anomalies both 
in the specific heat and the macroscopic susceptibility at low temperatures down to 2K.   
Figure 4 shows typical muon relaxation curves under LF up to 3900Oe.   
In 10K, entire muon relaxation is driven by the quasi static nuclear spin fluctuation, 
because an initial part of the relaxation decays with Gaussian-like function, 
and hence the relaxation is completely suppressed by a weak LF of a few tens Oe.   

In 0.33K, the relaxation is divided into the two components.   
One is the quasi static nuclear spin part, which has the Gaussian form, 
and is suppressed by a weak LF of 20Oe.   
The other is a dynamically-fluctuating electronic part, which has the Lorentzian form, 
and survives relatively high a field up to 3900Oe.   
The relaxation rate $\lambda$ was obtained by fitting the exponential function
to the relaxation curves under LF above 20Oe.
The fraction of the latter components is around one half of the entire asymmetry, 
indicating that, roughly speaking, a half of the muons are not affected by any electron spins.

\begin{figure}
\includegraphics*[trim=1cm 11cm 1cm 2cm, clip, scale=0.5]{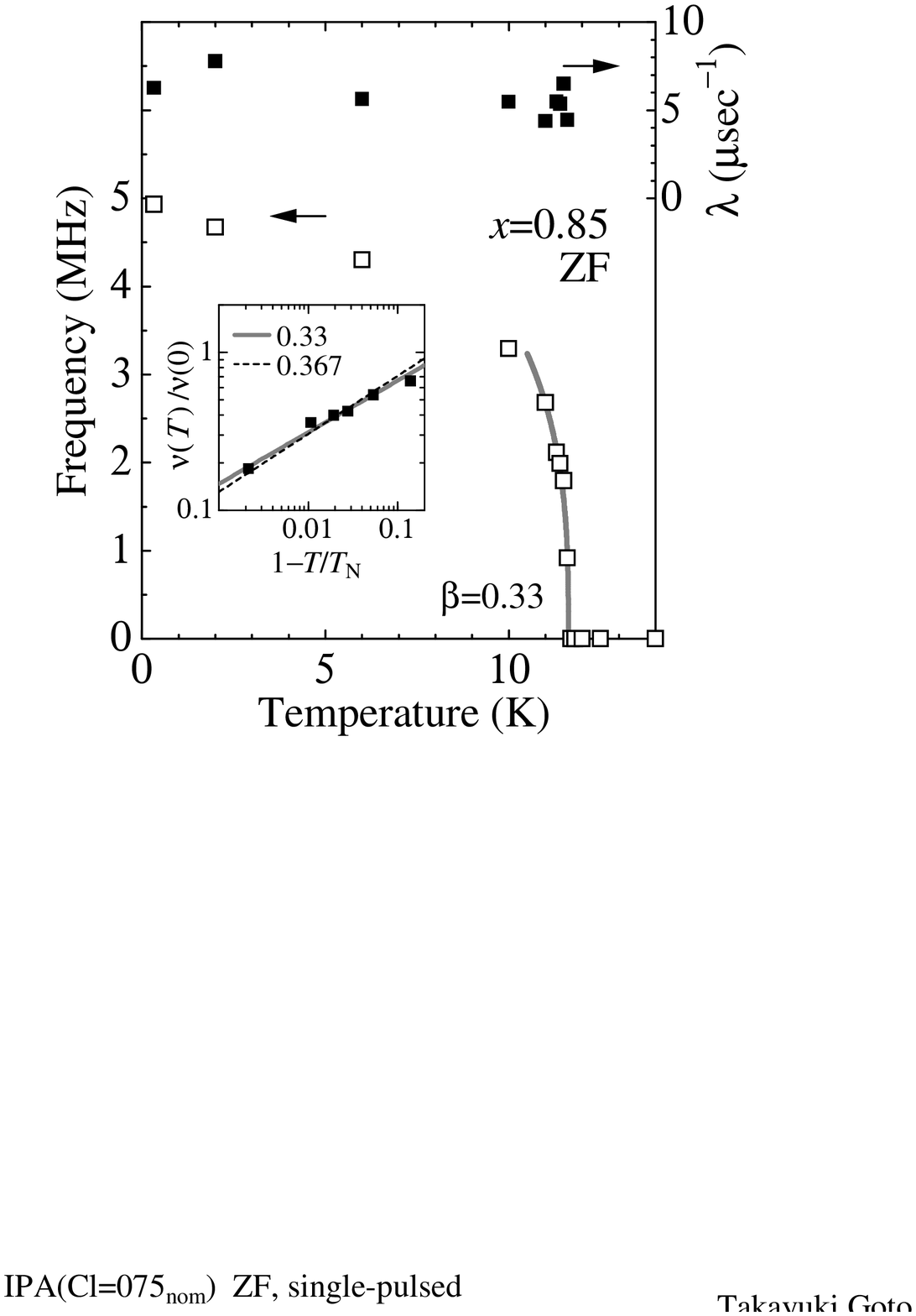}
\caption{
Temperature dependence of the rotation frequency $\nu$ (open square)
and relaxation rate $\lambda$ (solid square).
The curve is the scaling function $(1-T/T_{\rm N})^\beta$, 
where $T_{\rm N}$ and $\beta$ is the N$\acute{\rm e}$el temperature 
and the critical exponent.   
The two parameters are obtained by the least square method to be 
$T_{\rm N}$=11.65K 
and $\beta$=0.33($\pm$0.02) ({\em inset}), 
which is close the value for 3D-Ising model.
A fitting line with the exponent
of 3D-Heisenberg model $\beta$=0.367 is also shown for comparison.
}
\end{figure}

The external field (LF) dependence of $\lambda$ 
shown at Fig.~5 
directly maps the Fourier component of the fluctuating field produced by 3$d$-spins.   
Assuming that the spin fluctuation of 3$d$-spins has a classical Lorentzian spectrum, 
we fitted the function 
\[
\lambda(H_{\rm LF})=
\frac{2\cdot(\gamma_\mu\delta H_{\rm loc})^2\tau_{\rm C}}
{(1+(\gamma_\mu H_{\rm LF}\tau_{\rm C})^2)}
\]
to the observed data
to obtain the mean fluctuation time and the fluctuation amplitude to be 
$\tau_{\rm C}$=0.44$\mu$sec and $\delta H_{\rm loc}$=55 Oe.   
The temperature dependence of $\lambda$ under LF of 100Oe, 
which is high enough for a complete suppression of the effect of nuclear spins, 
is shown in Fig. 6.   
A significant increase that looks to be diverging at the zero temperature is observed below 5K.

Here we mention that in both the two samples with and without a static order, 
approximately half of the muons are relaxed by the local field produced by 3$d$-spins.   
This means that the other half only feel the local field
produced by nuclear spins, fluctuation of which is quasi static,
or by the 3$d$-spins fluctuating with frequency out of the muon time window.
We will discuss this point in the following section.


\begin{figure}
\includegraphics*[trim=2cm 4cm 1cm 2cm, clip, scale=0.4]{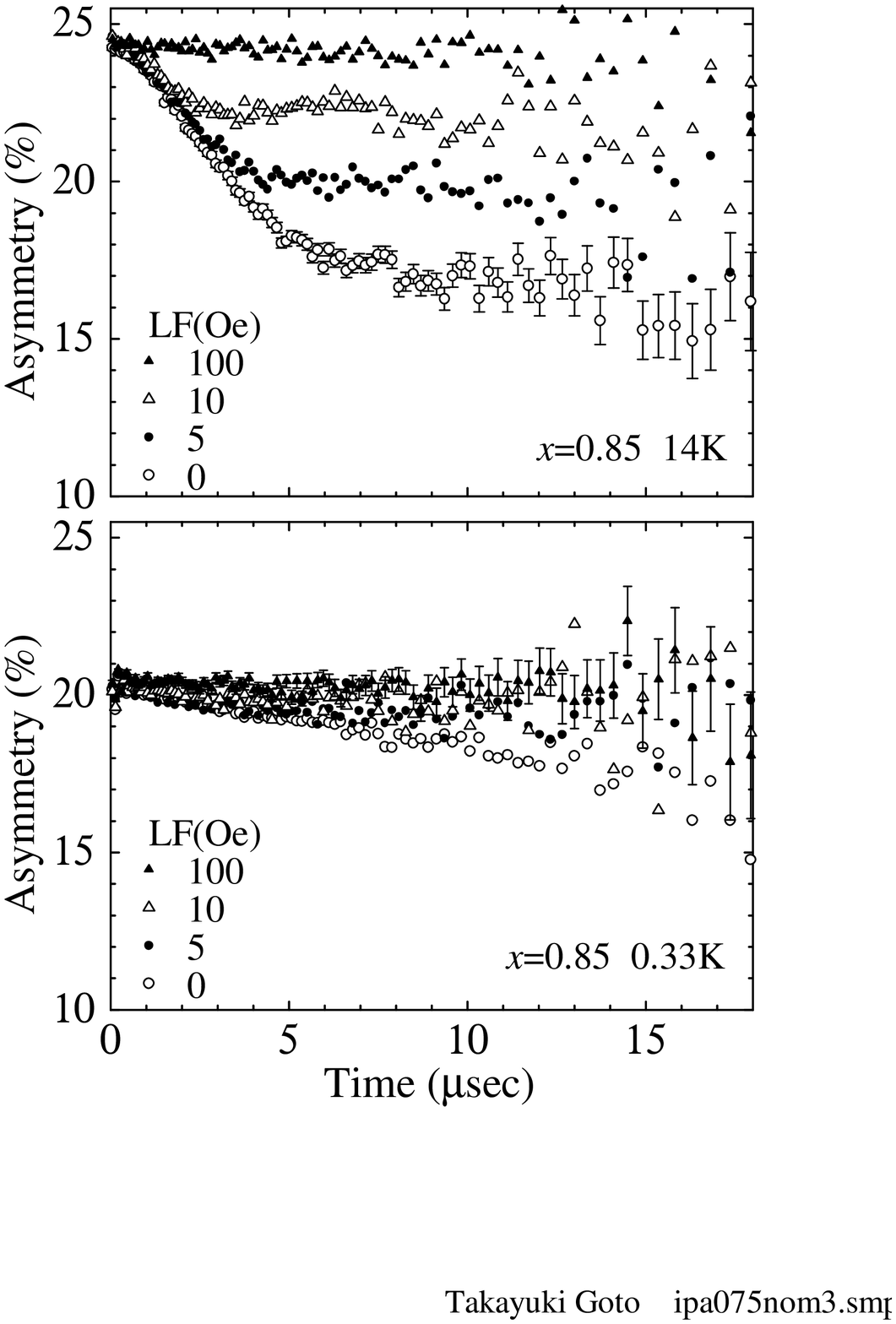}
\caption{
Muon relaxation spectra of the sample $x$=0.85 at paramagnetic state (upper panel, 14K) 
and ordered state (lower panel, 0.33K) under various longitudinal fields (LF) up to 100Oe.
}
\end{figure}

\section{Discussion}

The muon spin rotation observed at low temperatures $T<T_{\rm N}$=11.65K indicates the 
existence of the long-range and static magnetic order in the sample of $x$=0.85.
Judging from the single Fourier component observed in the rotational profile, the spin structure 
is considered to be simple antiferromagnetic.   
This result microscopically demonstrates that the solid solution of the 
two spin gap systems has a magnetic ground state.
The critical exponent of the order parameter $\beta$=0.33 agrees with the theoretical value of 
3D-Ising model~\cite{3D_Ising_exponent_experiment,3D_Ising_exponent_Monte_Carlo,3D_XY_critical_exponent_theory1,3D_XY_critical_exponent_text}.
The appearance of the three dimensional exponent is explained as
that the phase transition in this low dimensional spin system is set off by weak 
inter-ladder interactions, the path of which runs three-dimensionally in the system.
The Ising-type universality class is 
consistent with the existence of the effective uniaxial anisotropy $D^*$
due to the formation of ferromagnetic dimers in IPA-CuCl$_3$, 
reported recently by Manaka\cite{ESR_D_term,ESR_D_term2}.

Next, we discuss the volume fraction of the magnetic order.
As stated above, half of the muons are not affected 
by the local field produced by 3$d$-spins in both the two samples.
This observation indicates the two possibilities.
One is that there are the two muon stopping sites.
The first site is, as in many other compounds, 
located near the negative ions of Cl$^-$ or Br$^-$, 
which are close to Cu-3$d$ spins, and bears a large local field.   
Actually, if we assume that the ordered moment is around 1 $\mu_{\rm B}$\cite{Kanada_H_NMR}, 
and that its interaction with muon is of a classical dipole-dipole type, 
the distance between the muon site and the Cu atom is
calculated from the observed local field of 360 Oe
to be 0.3-0.37 nm, which is comparable with that between Cl and Cu atoms 0.22-0.23 nm,
supporting our idea.
The second site may be located, for example, 
around or inside the anion molecule (CH$_3$)$_2$CHNH$_3$, which is far from 3$d$-spins,
and its local field should be very small.
By considering this second site, we can understand
why the half of muons are not affected by the magnetic field produced by 3$d$-spins 
even though the volume fraction of the magnetic order is nearly unity.
The other possibility is that half of the
muons only experience electronic moments fluctuating outside the muon time window.
We emphasize here that
the possibility of the phase separation in samples is excluded,
because the sample quality is assured both by the macroscopic susceptibility
containing no spurious Curie-term,
and by the sharpness in the phase transition observed by the specific heat and
by the present muon experiment (Fig. 2).   
Furthermore, if half the volume of the sample $x$=0.85 is contaminated 
by other concentration which do not shows magnetic order ($x\geq$0.87), 
the half of the contaminating region is expected to show an appreciable 
relaxation due to the dynamical spin fluctuation as is observed in
the sample $x$=0.95.
This contradicts to the observed fact.

\begin{figure}
\includegraphics*[trim=1cm 4cm 0.7cm 2cm, clip, scale=0.4]{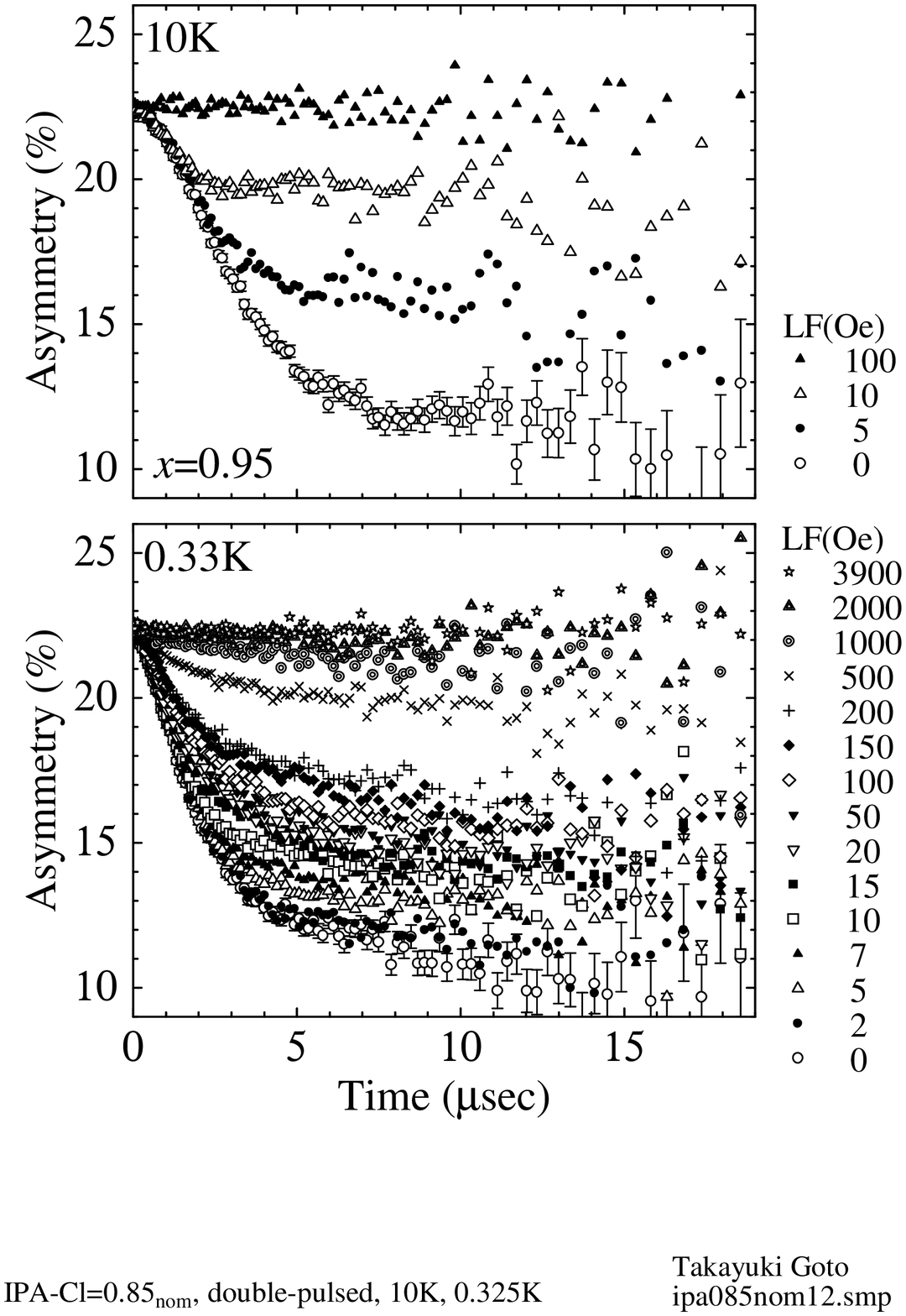}
\caption{
Muon relaxation spectra of the sample $x$=0.95 at 10K (upper panel) 
and 0.33K (lower panel) under various longitudinal fields (LF) up to 3900Oe.
}
\end{figure}

\begin{figure}
\includegraphics*[trim=1cm 11cm 1cm 1.2cm, clip, scale=0.5]{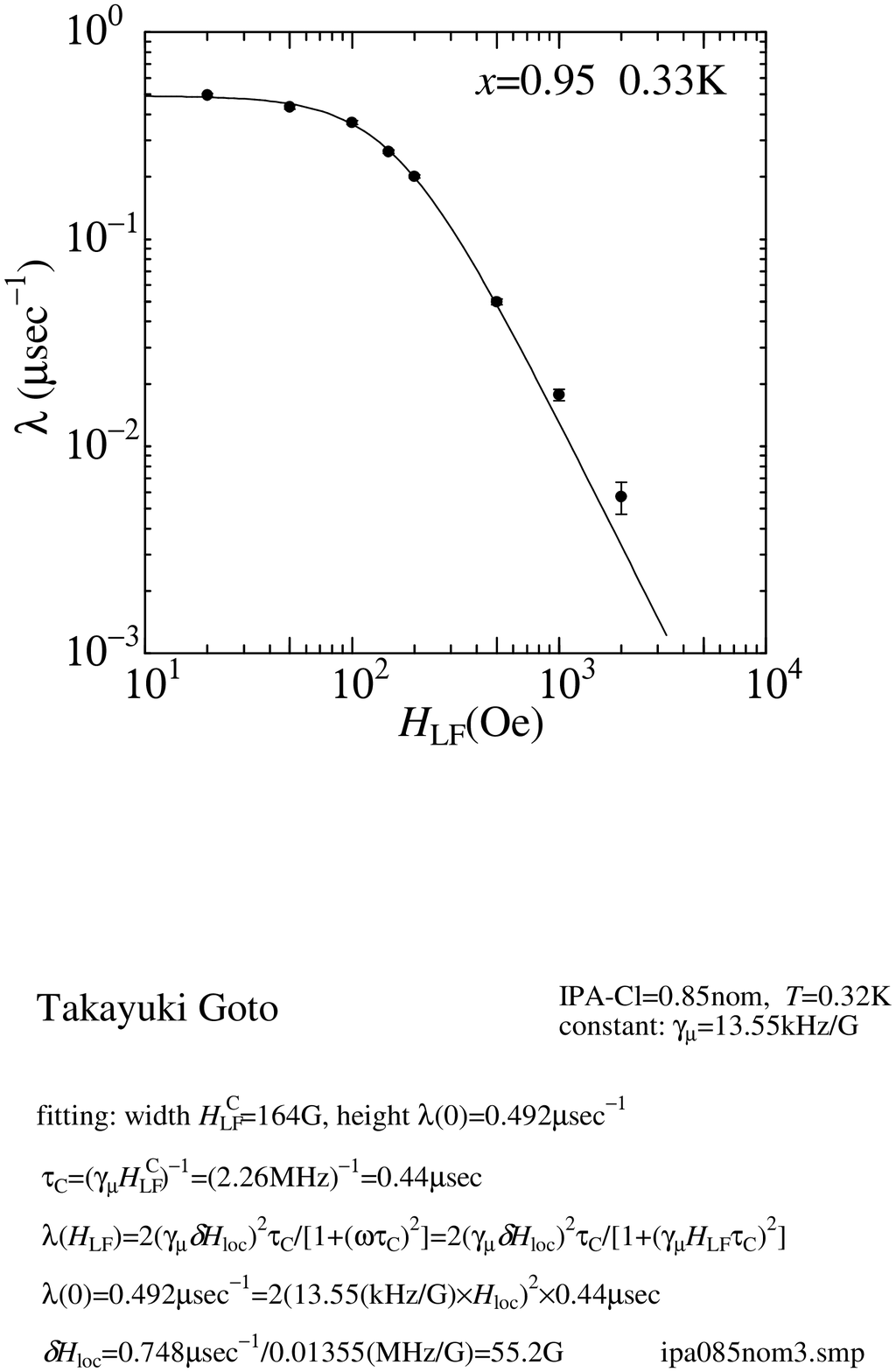}
\caption{
Magnetic-field dependence of the muon relaxation rate $\lambda$ in the sample $x$=0.95 
at the temperature 0.33K and 
under various longitudinal magnetic fields (LF) up to 3900Oe.   
The curve is the Lorentz function explained in the text.
}
\end{figure}

The profile of the relaxation curves in the paramagnetic state does not change from 
one at high temperatures until the temperature reaches at $T_{\rm N}$.   
This means that the temperature region where the critical slowing down of 3$d$-spins, if exists, starts and its characteristic frequency gets into the muon time window ($\nu<10^{-11}$sec$^{-1}$)
 must be extremely narrow, that is, $T-T_{\rm N}<$0.1K.  
On the other hand, the sample of $x$=0.95 shows a diverging behavior of $\lambda$
 toward the absolute zero, indicating that a critical slowing down of the spin 
fluctuation takes place in rather a wide range of temperature, though there is no long range order 
at low temperature down to 0.33K.
This sample of $x$=0.95 has been indicated to be gapped by 
the results of macroscopic measurements 
of Manaka' s group~\cite{chi_and_C_on_mixed_system_Br_Cl,M_H_on_mixed_system_Br_Cl}
and also of the present study. 
Nevertheless, the present muon observation clearly demonstrates that the system has 
the magnetic instability in the ground state.   
We can immediately identify the dominant q-part of this instability to be antiferromagnetic like, 
because the uniform susceptibility vanishes at low temperatures.   
In regard to the origin of the magnetic instability, we propose the two possibilities.

One is the precursor effect to 
the Bose-glass phase at the absolute zero~\cite{Fisher_Bose_glass}.
In solid-solution of the two spin-gap systems (Tl,K)CuCl$_3$, an anomalous enhancement 
of $\lambda$ as well as NMR-$T_1$ have been reported and the possibility 
of the Bose-glass phase is suggested.\cite{TlK_Suzuki_muSR,TlK_Fujiwara_NMR}
The other possible origin is 
the fluctuation effect around the quantum critical point at $x$=0.87, 
which divides the area into the antiferromagnetic and paramagnetic regions. 
In the samples $x$=0.95, though the static order does not occur, the antiferromagnetic
component of $\chi(\mbox{\boldmath $q$}\simeq \mbox{\boldmath $Q$}_{\rm AF},\omega)$ 
may be highly enhanced.
This is analogous to the case of high-$T_{\rm C}$ cuprates of the superconducting phase,
where a pronounced antiferromagnetic
fluctuation in 3$d$-spins persists, even though there is 
no magnetic order at finite temperatures.~\cite{Kitaoka_LSCO,Birgeneau_neutron_LSCO}
Finally, we do not deny a possibility of the long-range order in this sample at a finite 
temperature much lower than the present study.   
The experiment in the temperature region of dilution-refrigerator is now on the progress.

\section{Summary}

By an intensive $\mu$SR study on the solid solution of two spin-gap systems 
(CH$_3$)$_2$CHNH$_3$-Cu(Cl$_x$Br$_{1-x}$) ($x$=0.85, 0.95), 
we have confirmed that the sample $x$=0.85 shows a long range magnetic 
order at $T_{\rm N}$=11.65K.   
The local field at muon site is 360Oe at the lowest temperature.   
The critical exponent estimated from the temperature dependence of 
the local field is $\beta$=0.33, 
which agrees with the theoretical value of 3D-Ising model.   
In the sample of $x$=0.95, which does not show any long range order, 
a significant critical slowing down of the electron spin fluctuation toward 
the absolute zero was observed.   
The characteristic correlation time and fluctuation amplitude of 
the 3$d$-spin fluctuation at 0.33K was obtained as 0.44$\mu$sec and 160Oe.

\begin{figure}
\includegraphics*[trim=1cm 9cm 1cm 2cm, clip, scale=0.5]{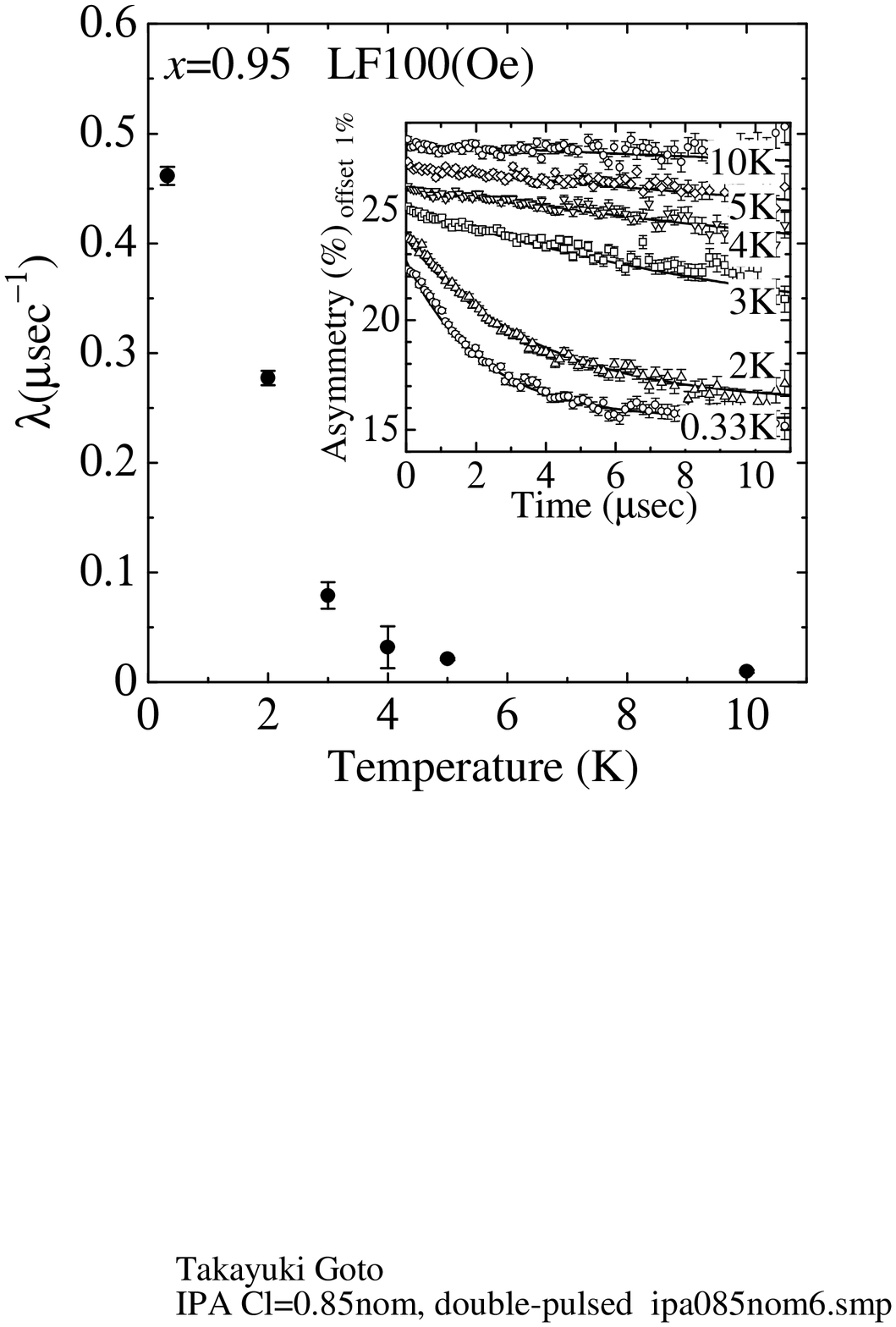}
\caption{
Temperature dependence of the muon relaxation rate $\lambda$ in the sample $x$=0.95 
under longitudinal field (LF) of 100Oe.   
The inset shows the profile of spectra at each temperatures.
Each spectrum other than that for 0.33K is shifted vertically by 1\%.   
Curves are fitted exponential functions from which $\lambda$ was obtained.
}
\end{figure}

\begin{acknowledgments}

The authors would like to show special thanks to H. Manaka for kind and important discussions 
on the experimental results 
and the sample syntheses.   
They also thank T. Masuda and A. Zheludev for a valuable discussion, 
and K. Noda for a kind assistance in measurements of the specific heat.   
This work has been supported by 
the Toray Science Foundation, 
Saneyoshi Scholarship Foundation, 
Kurata Memorial Hitachi Science and Technology Foundation,
the Joint Programs and Grant-in-Aid for the JSPS,  
and 
the Grant-in-Aid for Scientific Research on Priority Area from Ministry of Education, 
Culture, Sports, Science, and Technology of Japan.

\end{acknowledgments}



\begin{thebibliography}{22}
\bibitem{is_Br_ladder_or_not}
The Br-system, which is isomorphic with Cl-system, had been reported 
to be an alternating chain 
from the temperature dependence of the macroscopic susceptibility.\cite{chi-T_in_Br_system}.
No determinate reports to judge whether or not it is have been made until now.
\end{thebibliography}
\end{document}